\documentclass[reprint,
superscriptaddress,
 amsmath,amssymb,
 aps,
prb,
]{revtex4-2}

\usepackage{graphicx}
\usepackage{svg}
\usepackage{dcolumn}
\usepackage{bm}
\usepackage[colorlinks=true, , citecolor=blue, linkcolor=blue, urlcolor=blue]{hyperref}

\usepackage{color}
\usepackage[normalem]{ulem}

\begin{document}

\preprint{APS/123-QED}

\title{Optimizing LZSM protocol for high-fidelity gates in open-system fluxonium}
\newcommand{\affA}{\affiliation{Grupo de Circuitos Cuánticos Bariloche, Div. Dispositivos y Sensores, Centro Atómico Bariloche-CNEA, 8400 San Carlos de Bariloche, Argentina.}}
\newcommand{\affB}{\affiliation{Instituto Balseiro (Universidad Nacional de Cuyo), Bariloche, Argentina.}}
\newcommand{\affC}{\affiliation{Centro At\'omico Bariloche and Instituto Balseiro, 8400 San Carlos de Bariloche, R\'io Negro, Argentina.}}
\newcommand{\affD}{\affiliation{Instituto de Nanociencia y Nanotecnolog\'{\i}a (INN), CONICET-CNEA, Argentina.}}

\author{Santiago Ferreyra}
\affC
\author{Valentín Reparaz}
\affA\affB
\author{María José Sánchez}
\affC\affD
\author{Leandro Tosi}
\affA\affB
\author{Daniel Domínguez}%
 \email{daniel.dominguez@ib.edu.ar}
\affC

\date{\today}

\begin{abstract}
Quantum gates based on resonant Rabi oscillations are inherently slow for small-frequency qubits. They are also prone to errors due to counter-rotating terms. However, when the anharmonicity is sufficiently high, as in the fluxonium architecture, alternative manipulation protocols can outperform standard resonant driving. In this work, we implement fast, high-fidelity quantum gates based on a one-period Landau-Zener-Stückelberg-Majorana (LZSM) driving protocol. We derive analytical expressions that simplify the exploration of the parameter space while accounting for the multi-level structure of the circuit. Furthermore, we analyze the role of leakage,  discussing strategies to mitigate it and identifying regimes in which it becomes the dominant source of error. Finally, to evaluate the impact of dissipation on gate fidelity, we develop a robust formalism suitable for analyzing the open-system performance of quantum gates in the strong driving regime.     
\end{abstract}

\maketitle


\section{\label{sec:intro}Introduction}
The coherent manipulation of quantum systems is typically realized with resonant drives, where the amplitude of the drive in the rotating frame determines the rotation frequency \cite{krantz_2019,kwon_2021}. Consequently, faster quantum gates require larger drive amplitudes, which inherently introduce errors associated with counter-rotating terms \cite{Bloch1940,Shirley1965,Rower_2024} and leakage into non-computational levels \cite{Fazio_1999,Motzoi_2009,Ferron2010,Gambetta_2011,Wood_2018}. These limitations are particularly detrimental for low-frequency qubits, where gate speed is fundamentally bounded by the qubit frequency, and in systems with low anharmonicity \cite{Campbell2020,zhang_2021}. A compelling alternative paradigm utilizes non-resonant driving. In this context, Landau-Zener-Stückelberg-Majorana (LZSM) interferometry offers a significant advantage, enabling fast, high-fidelity rotations via single- or few-period sinusoidal pulses \cite{Campbell2020,Caceres2023,Ryzhov_2024}. 

Originally proposed for the spectroscopy of quantum systems using large-amplitude drives \cite{Shevchenko_2010,ivakhnenko_2023}, LZSM protocols have found an ideal testbed in superconducting qubits due to their unprecedented level of control and tunability \cite{ivakhnenko_2023,Oliver_2005,Oliver_2009,Sillanpaa_2006,Ferron_2012,Ferron_2016,Reparaz2025}. Recently, LZSM gates were successfully demonstrated in a low-frequency qubit \cite{Campbell2020}, and similar concepts have been applied to two-qubit gates featuring small anti-crossings \cite{weiss_2022}. 

In LZSM gates, one must navigate a parameter space defined by the drive frequency $\omega_d$ and amplitude $A_e$. The total gate time is given by $2\pi/{\omega_d}+t_{\text{idles}}$, which includes idle times necessary for phase correction. Optimization requires identifying parameter pairs $(A_e, \omega_d)$ that simultaneously maximize fidelity and minimize gate time. While valuable analytical tools have been developed recently to simplify this experimental tuning \cite{Caceres2023,Ryzhov_2024}, these works are restricted to pure two-level systems and consider only unitary evolution. 


In this work, we develop a theoretical framework for the optimization of LZSM quantum gates that explicitly incorporates the multilevel structure of realistic devices. Although our analysis focuses on a fluxonium superconducting circuit \cite{manucharyan_2009,nguyen_2019,Nguyen2022,bao_2022,somoroff_2023}, the methodology is applicable to other qubit architectures. We analyze leakage errors and propose a targeted mitigation strategy. Crucially, we also present a theoretical method for calculating gate fidelity in the presence of dissipation. Standard models often neglects the dependence of the system-bath coupling on the drive amplitude, an approximation that breaks down in the strong-driving regime relevant to LZSM protocols \cite{kohler_1997,grifoni_1998,hone_2009,hausinger_2010,Ferron_2012,Ferron_2016}.
Our model captures this dependence, allowing us to establish rigorous bounds for gate fidelity based on qubit lifetime and bath temperature. 

The manuscript is organized as follows: In Sec. II, we analyze the LZSM protocol restricted to a two-level system dynamics. In Sec. III, we introduce the fluxonium circuit and in Sec. IV we detail the dynamics of the full multi-level system, discussing effective models and analyzing leakage. In Sec. V, we present our framework for calculating the open-system fidelity. Finally, in Sec. VI we discuss our results.

\section{\label{sec:analytical_model}Two-level system dynamics}
\begin{figure}[!t]
\includegraphics[width=1.0\linewidth]{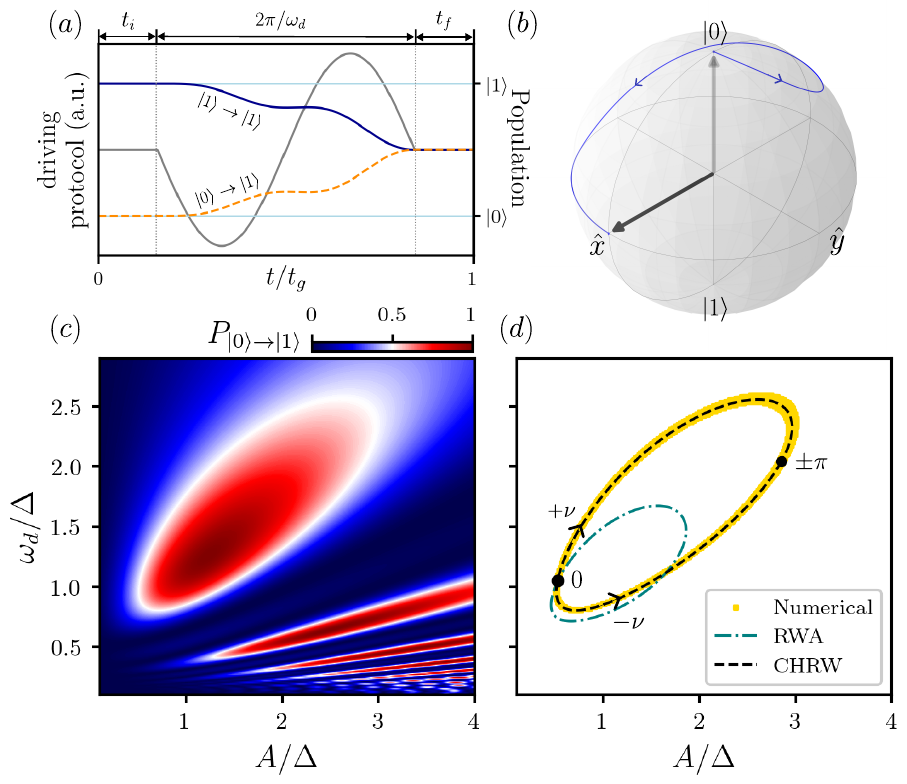}
\caption{\label{fig:TLS_protocolo_P01}
\textbf{LZSM dynamics in an ideal two-level system.} (a) Driving protocol (left axis, grey line), consisting of two idle times $t_i$, $t_f,$ and one single-period sinusoidal pulse. The driving parameters correspond to $(A/\Delta, \omega_d/\Delta)=(1.02, 1.78)$. Blue and orange lines show the evolution of the population over time (right axis). (b) Evolution on the Bloch sphere of the initial state $|0\rangle$ (grey arrow), onto the final state $(|0\rangle + |1\rangle)/\sqrt 2$ (dark arrow). (c) Transition probability as a function of driving parameters. (d) Region where $P_{01}=1/2$, compared with the analytical curve obtained from the standard RWA (dash-dotted line), and the improved CHRW approximation (dashed line). The CHRW curve is parametrized by $\nu\in[-\pi,\pi)$, from Eq. \eqref{eq:nu_equation}.}
\end{figure}

For the implementation of quantum gates we start discussing the dynamics of a transversely driven two-level system (TLS) with Hamiltonian
\begin{equation}
    H_\text{TLS}(t) = -\frac{\Delta}{2}\hat{\sigma}_z-\frac{A}{2}\sin(\omega_d t)\,\hat{\sigma}_x,
    \label{eq:H_TLS}
\end{equation}
where $\Delta$ is the qubit frequency 
and $A$ the 
driving amplitude. We take $\hbar=1$ and $\hat{\sigma}_x$, $\hat{\sigma}_y$, $\hat{\sigma}_z$ are Pauli matrices. 
We analyze the case of a single period $T=2\pi/\omega_d$ driving protocol \cite{Campbell2020, Caceres2023}, with 
frequency $\omega_d$ (see Fig. \ref{fig:TLS_protocolo_P01}(a)). We do not restrict the driving amplitude to small values $A/\Delta$ as in the case of resonant Rabi-like many-periods microwave control. Under these conditions, population transfer occurs via non-adiabatic Landau-Zener-Stückelberg-Majorana (LZSM) transitions \cite{ivakhnenko_2023}. Additional idling times are introduced to correct the phase in order to implement a desired gate \cite{Campbell2020,Caceres2023}. Taking a $Y(\pi/2)$ gate as a example, in Fig. \ref{fig:TLS_protocolo_P01}(b) the evolution of the initial state $|0\rangle$ in the Bloch sphere is shown for the full dynamics of the protocol indicated in (a). 

With the scope of implementing $\pi/2$ gates, and proceeding as in an experimental calibration \cite{Campbell2020}, we first explore the $(A/\Delta, \omega_d/\Delta)$ parameter space and plot in Fig. \ref{fig:TLS_protocolo_P01}(c) the probability of finding the system in the state $|1\rangle$ at the end of the protocol applied to the initial state $|0\rangle$. The numerical results are obtained from integrating the Schrödinger equation. The color code is chosen such that the white contour indicates the midpoint where the state is pointing to the equator i.e. $P_{01}=1/2$. The corresponding parameters would allow to implement $\pi/2$ rotations, and therefore a gate like $Y(\pi/2)$ after correcting eventually the phase with a rotation around $Z$.

Precise analytical solutions in the ideal TLS case  \cite{Yan2015, Caceres2023} allow to estimate the driving parameters for a desired unitary. This is very handful since it avoids the tedious experimental work of exploring all the parameter space \cite{Campbell2020}. The method is based on a counter-rotating hybridized rotating wave approximation (CHRW) \cite{Yan2015}, which correctly predicts the existence of high-fidelity $\pi/2$ rotations for a wide range of driving parameters, with $\omega_d$ ranging from $0.8\Delta$ to $2.5\Delta$, and amplitudes as large as $3\Delta$. This parameter region is given by the solutions to the following non-linear equation,
\begin{equation}
    \sin^2(\alpha)\sin^2(\beta)=\frac12,
    \label{eq:analytical_equation_P01=0.5}
\end{equation}
where $(\alpha, \beta)$ depend on the normalized
driving parameters $(A/\Delta,\omega_d/\Delta)$ (see details in App. \ref{A:CHRW}). The set of solutions to Eq. \eqref{eq:analytical_equation_P01=0.5} is shown in Fig. \ref{fig:TLS_protocolo_P01}(d), forming a closed curve in the parameter space. The CHRW approximation successfully reproduces the numerical results as compared to a simple rotating-wave approximation (RWA). Every point on this analytical curve can be parametrized in terms of a phase $\nu\in[-\pi,\pi)$, 
\begin{equation}
    \frac{1}{\sqrt2}e^{i\nu}=\cos(\alpha)+i\sin(\alpha)\cos(\beta),
    \label{eq:nu_equation}
\end{equation}
that can be considered of as the tuning knob of this scheme. In what follows we show that in the case of the fluxonium circuit, the  optimal value of $\nu$ will depend non-trivially on the multi-level structure of the full system and on the coupling with the environment. 

\section{\label{sec:fluxonium}Fluxonium Circuit}

\begin{figure}[!t]
\includegraphics[width=1.0\linewidth]{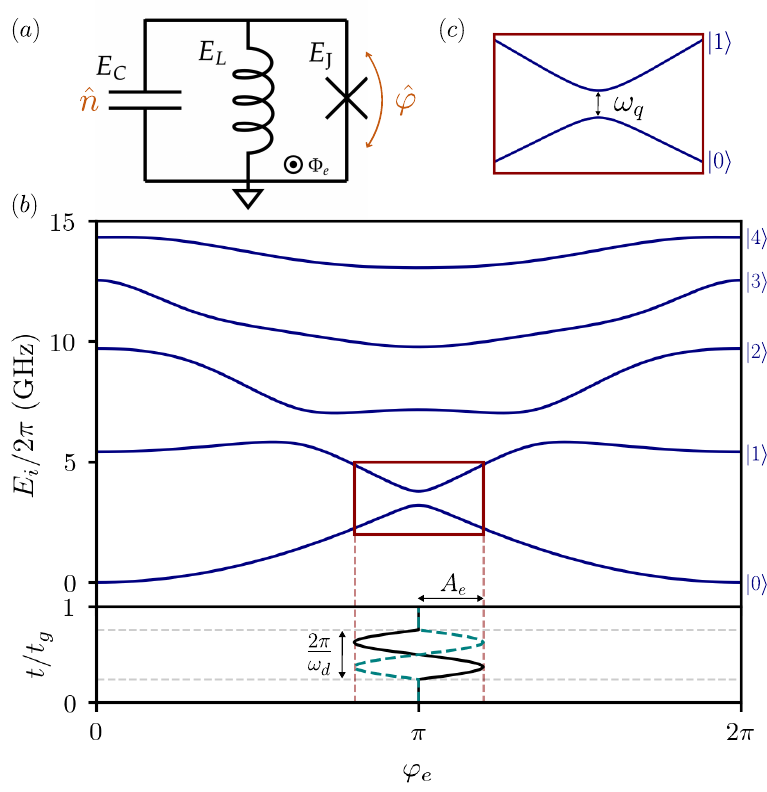}
\caption{\label{fig:fluxonium_energies_protocol}(a) Superconducting fluxonium circuit, consisting of a capacitor, an inductor and a Josephson junction in parallel. An external magnetic flux is threaded into the loop. (b) Upper panel: Low-energy spectrum of $H_0$ as a function of the control variable $\varphi_e=2\pi\Phi_e/\Phi_0$, in the case of $\{ E_J, E_C, E_L\} = \{4,1,1 \}\times2\pi$ GHz. Lower panel: Driving protocol $\varphi_e(t)$ including idle times, for the two possible sign choices for the amplitude $A_e$ (solid and dashed lines, respectively). (c) Avoided crossing between states $|0\rangle$ and $|1\rangle$ near the sweet-spot $\varphi_e=\pi$, suitable for LZSM control.}
\end{figure}

The TLS model of Eq. \eqref{eq:H_TLS} is the most simple approximation to any realistic superconducting circuit. We consider in the following the fluxonium architecture \cite{manucharyan_2009}. The circuit (see Fig. \ref{fig:fluxonium_energies_protocol}(a)) consists of an inductor and a Josephson junction forming a loop, shunted by a capacitor. The Hamiltonian of the system can be written as \cite{manucharyan_2009,nguyen_2019,Nguyen2022}
\begin{equation}
    H_0 = 4 E_C \hat{n}^2 + \frac12 E_L \left( \hat{\varphi} - \varphi_e\right)^2 - E_J\cos(\hat{\varphi}),
    \label{eq:H_fluxonium}
\end{equation}
where $\hat{n}$ and $\hat{\varphi}$ are the circuit's conjugate degrees of freedom $[\hat{\varphi},\hat{n}]=i$, describing the number of excess Cooper pairs across the junction and the junction's superconducting phase difference, respectively. The external magnetic flux applied into the loop is $\varphi_e=\Phi_e/\phi_0$, expressed in units of the reduced flux quantum, $\phi_0=\Phi_0/2\pi$. Also, $E_C=e^2/2C$, $E_L=\phi_0^2/L$, and $E_J=\phi_0I_c$, are the charging, the inductive and the Josephson energy respectively, in terms of the capacitance $C$, the inductance $L$ and the critical current $I_c$. The fluxonium regime is characterized by energy scales that satisfy roughly $ 0.1\leq\left\{ E_C/E_J,\,E_L/E_J \right\}\leq0.5 $ \cite{manucharyan_2009,nguyen_2019,Nguyen2022,bao_2022,somoroff_2023}, which define a double-well potential at the sweet-spot $\varphi_e=\pi$ with a small tunneling probability across the well. The fluxonium is characterized by relatively flat bands with respect to $\varphi_e$, which enhances the coherence, low qubit frequencies $\omega_{10}/2\pi\sim500$ MHz, and a large anharmonicity at the sweet-spot (see Fig. \ref{fig:fluxonium_energies_protocol}(b)). A zoom-in of the region of the spectrum that can be described by  Eq. \eqref{eq:H_TLS} is shown in  Fig. \ref{fig:fluxonium_energies_protocol}(c), presenting an avoided-crossing around $\varphi_e=\pi$, suitable for LZSM control.

We consider the drive through a fast antenna which produces a time-dependent magnetic flux, $\varphi_e \rightarrow\varphi_e(t)=\pi+\delta\varphi_e(t)$. The Hamiltonian is effectively given by \cite{Reparaz2025}
\begin{equation}
    H(t) = H_0 - E_L\delta\varphi_e(t)\hat{\varphi},
    \label{eq:driven_hamiltonian}
\end{equation}
which can be written in terms of the static basis of $H_0$ at $\varphi_e=\pi$,
\begin{equation}
    H(t) = \sum_i E_i |i\rangle\langle i| - E_L \delta\varphi_e(t) \sum_{ij}\varphi_{ij}|i\rangle\langle j|,
    \label{eq:driven_hamiltonian_eigenbasis}
\end{equation}
where the states $\{|i\rangle\}$ satisfy $H_0|i\rangle= E_i|i\rangle$, with $i=0,1,\dots$, and $\varphi_{ij} \equiv \langle i|\hat{\varphi}|j\rangle$. In the following we discuss the dynamics of the full circuit under the single-period sinusoidal pulse drive of amplitude $A_e$ and frequency $\omega_d$, where $\delta\varphi_e(t)=A_e\sin\omega_dt$, and complemented with pre and post idlings where $\delta\varphi_e(t)=0$. The full protocol is schematized in the lower panel of Fig. \ref{fig:fluxonium_energies_protocol}(b). We show next how the LZSM gate scheme can be applied to a fluxonium circuit with realistic parameter choices. 

\section{\label{sec:coherent-error}
Coherent dynamics and gate fidelity}

Following  the analysis of Ref. \cite{Nguyen2022}, we first discuss the case of a moderate architecture $\{ E_J, E_C, E_L\} = \{4,1,1 \}\times2\pi$ GHz, which corresponds to a compact and reliable qubit with a low transition frequency, $\omega_{10}/2\pi=0.58$ GHz, where $\omega_{ij}\equiv E_i - E_j$. The second excited state is separated by $\omega_{21}/2\pi=3.39$ GHz, giving a large anharmonicty $\alpha=(\omega_{21}-\omega_{10})/2\pi=2.81$ GHz.

\begin{figure*}[!t]
\includegraphics[width=.9\linewidth]{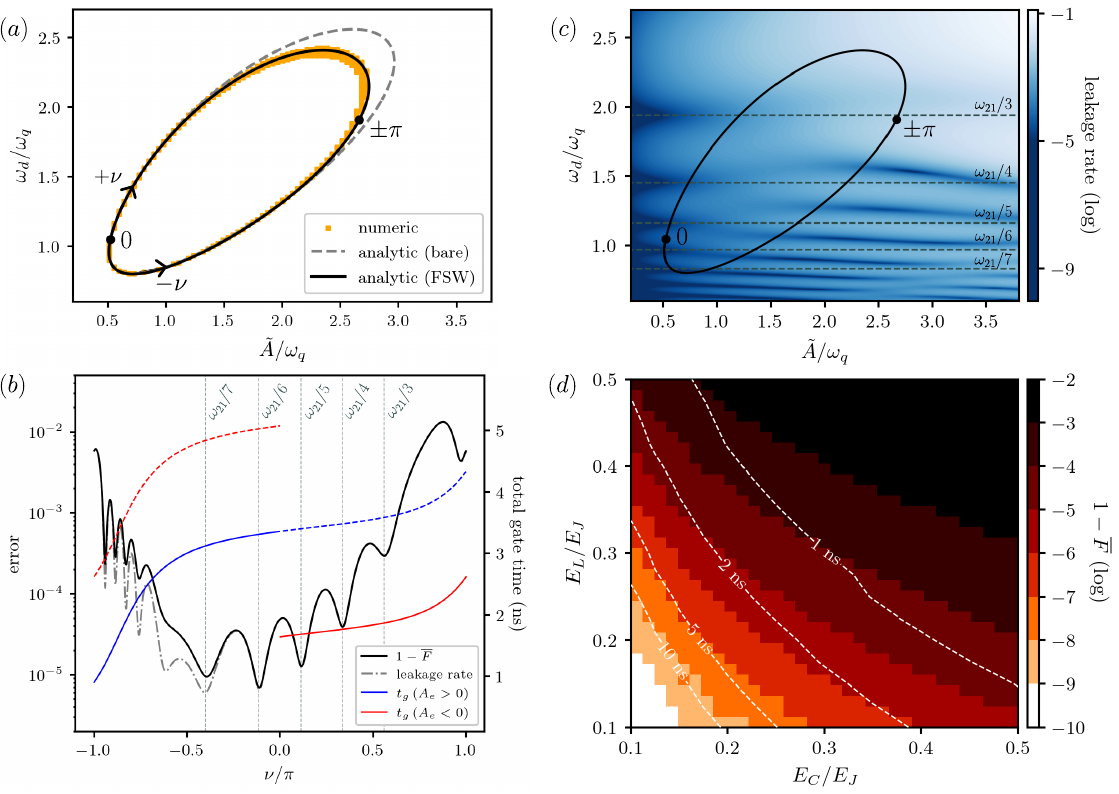}
\caption{\label{fig:fxnm_closed_errors}\textbf{Coherent error sources in the multi-level fluxonium circuit.} (a) Region where $P_{01}=1/2$ (orange) as a function of the driving parameters, compared with the analytical curves calculated from bare truncation (dashed line) and FSW effective two-level system (solid line). The architecture in (a, b, c) corresponds to $\{ E_J, E_C, E_L\} = \{4,1,1 \}\times2\pi$ GHz. (b) $Y(\pi/2)$ gate error calculated on the FSW analytical curve (dark line, left axis). The right axis shows the total gate time $t_g$ for both positive $A_e$ (blue) and negative $A_e$ (red). The solid segments correspond to the choice of minimal time for each value of $\nu$. (c) Leakage rate $L(t_g)$ in log scale, as a function of the driving parameters. The solid line shows the analytical FSW curve, and the dashed horizontal lines indicate the privileged $\omega_{21}/n$ frequencies, also indicated as vertical dotted lines in (b). (d) Minimum gate error over architecture space within the fluxonium regime, taking the optimal value from $\nu\geq0$ for faster gate times. Dashed contour lines show the total $t_g$, including idling time, with fixed $E_J/2\pi=4$ GHz.}
\end{figure*}

\subsection*{The $P_{01}=1/2$ contour}
From the full Hamiltonian in Eq. \eqref{eq:H_fluxonium}, we keep the lowest $5$ levels, which we have verified to be enough to describe the full multi-level dynamics for this circuit \cite{Reparaz2025}. We then numerically integrate  the Schrödinger equation for the Hamiltonian in Eq. \eqref{eq:driven_hamiltonian} applying a 4th order Trotter-Susuki algorithm \cite{Hatano2005}, which provides a $5\times 5$ unitary time-evolution operator $U(t)$. At the end of the protocol, for each amplitude and frequency of the pulse we calculate the transition probability $P_{01}(A_e,\omega_d)=|\langle 0|U(t_g)|1\rangle|^2$. Here $t_g=t_i+t_f+{2\pi}/{\omega_d}$ with two idles of duration $t_i$ and $t_f$ before and after the sinusoidal pulse used to correct the phase for a desired gate \cite{Campbell2020,Caceres2023}. In Fig. \ref{fig:fxnm_closed_errors}(a) we present the region where $P_{01}=1/2$. As in the case of a TLS shown in Figs. \ref{fig:TLS_protocolo_P01}(c,d), the contour is still an ellipse-like closed curve. In order to apply the analytical tools as in Sec. \ref{sec:analytical_model}, we need to obtain the effective  two-level system dynamics. A straightforward approximation is obtained by a bare truncation into the lowest two levels of the Hamiltonian \eqref{eq:driven_hamiltonian_eigenbasis}, which yields
\begin{equation}
    H_q^\text{bare}(t) = - \frac{\omega_q}{2}\hat{\sigma}_z - \frac{\tilde{A}}{2}\sin(\omega_d t)\,\hat{\sigma}_x,
    \label{eq:effective_Hq_bare}
\end{equation}
where $\omega_q \equiv \omega_{10}$ is the qubit frequency in the absence of driving and $\tilde{A} \equiv 2A_e E_L\varphi_{01}$ is the effective amplitude inside the qubit's subspace. We are making use here of the fluxonium selection rules at the sweet-spot which constrain $\varphi_{ij}\neq 0$ only for $i+j= 1\,(\text{mod 2})$, thus $\hat{\varphi} \rightarrow \varphi_{01}\hat{\sigma}_x$.

In Fig. \ref{fig:fxnm_closed_errors}(a) we plot the analytical curve obtained using the CHRW approximation with the truncated Hamiltonian of Eq. \eqref{eq:effective_Hq_bare}. The curve (dashed-line) is generated  by sweeping over $\nu\in[-\pi,\pi)$, and for each value solving Eqs. \eqref{eq:analytical_equation_P01=0.5} and \eqref{eq:nu_equation}.  We note that this curve deviates from the numerical result at large amplitudes. As we show next, with good accuracy, this effect is due to a renormalization of the qubit frequency in terms of the drive parameters, $\tilde{\omega}_q(\tilde{A},\omega_d)$. 

We start from the Hamiltonian in Eq. \eqref{eq:driven_hamiltonian_eigenbasis} and write it in a Floquet frame. By performing a Schrieffer-Wolf transformation (see details in App. \ref{A:FSW-approximation}), we obtain a perturbative description of the effective dynamics inside the computational basis which accounts for the presence of higher energy levels \cite{Reparaz2025}. Up to leading order $\tilde{A}^2$, the  effective Floquet-Schrieffer-Wolf (FSW) Hamiltonian can be written as,
\begin{equation}
    H_q^\text{FSW}(t) = -\frac{\tilde{\omega}_q(\tilde{A}, \omega_d)}{2}\hat{\sigma}_z - \frac{\tilde{A}}{2}\sin(\omega_d t)\,\hat{\sigma}_x
    \label{eq:effective_H_FSW},
\end{equation}
where
\begin{equation}
    \begin{aligned}
        &\tilde{\omega}_q(\tilde{A}, \omega_d) = \tilde{E}_1 - \tilde{E}_0\\
        &= \omega_q + \frac{\tilde{A}^2}{2}\sum_{j\geq 2} \left[\varphi_{j0}^2 \frac{\omega_{j0}}{\omega_{j0}^2 - \omega_d^2} - \varphi_{j1}^2 \frac{\omega_{j1}}{\omega_{j1}^2 - \omega_d^2} \right].
    \end{aligned}
    \label{eq:tilde_wq}
\end{equation}
As shown in Fig. \ref{fig:fxnm_closed_errors}(a), a major improvement is obtained by solving the analytical curve using the effective FSW Hamiltonian from Eq. \eqref{eq:effective_H_FSW}. The renormalized curve completely fits now the numerical contour. Although higher order corrections to Eq. \eqref{eq:tilde_wq} can be calculated by considering multi-photon processes in the extended Sambe space and  amplitude corrections to the quasienergies,  we  have  verified numerically that the present description is sufficient 
for the studied  regime.

\subsection*{Gate fidelity}

The FSW analytical curve plotted in Fig. \ref{fig:fxnm_closed_errors}(a) allows to identify the parameter set $(A_e,\omega_d)$ where  Y$(\pi/2)$ gates can be implemented in the fluxonium, taking into account its multilevel structure. However, these gates will not be perfect, since  pulses with large amplitudes $A_e$ can produce transitions to higher energy states beyond the computational space, inducing errors. By calculating the gate error in each point of the $P_{01}=1/2$ curve, we can find the optimal parameters $(A_e,\omega_d)$ for gate implementation within the LZSM protocol. We measure the gate error, or infidelity, as $1-\bar{F}$, where $\bar{F}$ is the average gate fidelity defined as \cite{pedersen_2007}

\begin{equation}
    \bar{F} = \frac{1}{d(d+1)}\left[ \Big|\text{Tr}\left(U^\dagger_q U_\text{target} \right)\Big|^2 + \text{Tr}\left(U_q^\dagger U_q\right)\right],
    \label{eq:def_fidelidad_closed}
\end{equation}
where $d$ is the dimension of the Hilbert space ($d=2$ for single qubit gates), $U_q$ is the time evolution operator inside the qubit's subspace extracted from the total unitary evolution of the multi-level system, $U_\text{total}(t)=U_q(t)U_\text{others}(t)$, and $U_\text{target}$ is the ideal target unitary, here $U_\text{target} = Y(\pi/2)$. In Fig. \ref{fig:fxnm_closed_errors}(b) we present the result of the gate error along the FSW analytical curve, parametrized by $\nu$. For each value, we set the driving amplitude and frequency by solving Eqs. \eqref{eq:analytical_equation_P01=0.5} and \eqref{eq:nu_equation}. Afterwards we compute   the idle times $t_i$ and $t_f$ given  in  Eq. \eqref{eq:idle_times_nu} of App. \ref{A:idles} and calculate $1-\bar{F}$ after the full evolution. First, we observe that the indfidelity is lower for small $\nu \sim 0$, which corresponds to smaller drive amplitudes. Second, the infidelity exhibits a sequence of sharp dips, at which the error is strongly suppressed. As we show below, both effects are linked to the leakage outside the computational basis.

We are also interested in minimizing the total gate time, and for this it is necessary to account for the idle times.  The simplest way to shorten the idling is by letting the amplitude take negative values, which is equivalent to implementing the same rotation but with opposite angle (see App. \ref{A:idles}). In Fig. \ref{fig:fxnm_closed_errors}(b) we show the total $t_g$ for both choices of the sign of $A_e$ (blue and red lines, respectively). Picking the optimal sign for each $\nu$, we see that much faster gates can be obtained for the case of $\nu\geq0$, and that a discrete jump of exactly a Larmor period $\tau_L=2\pi/\omega_q$ appears at $\nu=0$. Comparing the error values obtained for $\nu<0$ and $\nu\geq0$, we conclude that the tradeoff is favorable for positive values of $\nu$ since it is possible to minimize $t_g$ while keeping high-fidelity gates. For these fluxonium parameters, the full analysis shows that the optimal operation point would be $\nu/\pi\approx0.1$, with an error on the order of $10^{-5}$ and a gate time of 1.92 ns.

\subsection*{Leakage errors}
To better understand the gate fidelity results of Fig. \ref{fig:fxnm_closed_errors}(b), we quantify the amount of probability being lost to higher energy states, by calculating the leakage rate \cite{Wood_2018},

\begin{equation}
    L(t) = 1 - \frac12\text{Tr}\left( U_q(t)^\dagger U_q(t)\right).
    \label{eq:leakage_def}
\end{equation} 

In Fig. \ref{fig:fxnm_closed_errors}(c) we show a color map of the leakage error $L(t_g)$ in log scale as a function of the driving parameters. The leakage mainly increases with the amplitude of the pulse, with a difference of 3 orders of magnitude between the $\nu=0$ and $\nu=\pm\pi$ regions. An interesting feature shows up in the frequency dependence: local minima appear even far from the resonant condition $\omega_d=\omega_q$. Near $\tilde{A}=0$, these minima appear exactly at $\omega_d^n = \omega_{21}/n$, for any integer $n\geq2$. To understand this, we first notice that driving induced transitions to higher energy states are determined by the matrix elements of the driving operator $\hat{\varphi}$. We find that the predominant matrix element outside the computational basis is $\varphi_{12}$, since at the sweet spot $\varphi_{02} = 0$. Therefore,  the main source of leakage comes from the $|1\rangle\to|2\rangle$ driving induced transition.  The probability amplitude of this transition is related to the Fourier transform of the driving pulse evaluated at $\omega_{21}$,
\begin{equation}
    \mathcal{FT}[\delta\varphi_e]
    \propto \frac{\tilde{A}\omega_d}{\omega_{21}^2-\omega_d^2}\sin\left(\pi\frac{\omega_{21}}{\omega_d}\right),
    \label{eq:pulse_spectrum}
\end{equation}
which vanishes when $\omega_{21} =n\omega_d$, for $n\geq2$. Therefore, for  $\omega^n_d=\omega_{21}/n$ the probability of transitioning to state $|2\rangle$ is minimized and leakage is suppressed. This means that driving at these privileged $\omega_d^n$ frequencies is preferred for optimal control. As the amplitude increases, these minima get distorted due to the renormalization of the states, but the present analysis remains still valid.

As shown in Fig. \ref{fig:fxnm_closed_errors}(b) almost all the infidelity curve is bounded by the leakage rate, and actually saturated for a large section of the curve, mainly $\nu\geq 0$, where $1-\bar{F} = L$. Further optimization can be done numerically (see App. \ref{A:fid-optimization}), where we obtain $1-\bar{F} = L$ for the full range of $\nu$. Given that the error is directly attributable to the leakage rate, the optimal fidelity is locally found at the $\nu$ values associated with the privileged $\omega_{21}/n$ frequencies. For these fluxonium parameters, the global minimum appears at $\nu/\pi\approx-0.1$, corresponding to $\omega_d =\omega_{21}/6$. However, as we discussed previously, the optimal fastest operation point is $\nu/\pi\approx0.1$, or $\omega_d=\omega_{21}/5$.

\subsection*{Architecture parameters}
One way to reduce the leakage rate and increase the gate fidelity is to change the architecture to obtain a lower qubit frequency and higher anharmonicity. In Fig. \ref{fig:fxnm_closed_errors}(d) we present the results for the minimum gate error (restricted to $\nu\geq0$ for faster gate time) as a function of $E_C/E_J$ and $E_L/E_J$, across the fluxonium regime. Total gate time is shown as contour lines. We find that for a fixed error threshold, an improvement in $t_g$ can be obtained by reducing $E_L/E_J$ and increasing $E_C/E_J$. For example, one can reach the threshold of $10^{-6}$ at $E_L/E_J=0.1$ and $E_C/E_J=0.375$, with a gate time of only 2 ns.

Notably, we found that there is a simple analytical expression that describes the way the leakage rate varies with the device's parameters (see details in App. \ref{A:leakage_analysis}). As a rule of thumb 
\begin{equation}
    1 - \bar{F} \sim \frac{1}{32}\left(\frac{\varphi_{21}}{\varphi_{10}}\right)^2\left(1 - \frac{\alpha}{\omega_{21}}\right)^4,
    \label{eq:analytic_fidelity_estimation}
\end{equation}
with $\alpha=\omega_{21}-\omega_q$. This equation provides an estimation of the order of magnitude for the gate error expected for a given architecture, which can be handful at the stage of designing the device.

As $E_C/E_J\to0$ or $E_L/E_J\to0$, the circuit approaches the heavy-fluxonium regime \cite{zhang_2021} and gate fidelity improves, with an error as low as $10^{-10}$. At this point, however, $\omega_q/2\pi$ approaches values as small as $10$ MHz, which has a Larmor period of 100 ns and therefore produces longer total gate times, limited essentially by the idles. 

\section{\label{sec:oqs-fidelity}Open System Gate Fidelity}

In the heavy-fluxonium regime, with coherent errors below $10^{-6}$, and $\omega_q/2\pi$ in the range of $\left[10-200\right]$ MHz, we can be certain that gate fidelity will be ultimately limited by decoherence processes and thermal fluctuations. These considerations are essential for finding an optimal regime in the architecture's parameter space, where the LZSM protocol may achieve an advantage over resonant driving protocols.

We consider an open quantum system approach based on the Floquet-Born-Markov master equation \cite{kohler_1997,grifoni_1998,hone_2009, hausinger_2010,Ferron_2012,Ferron_2016},
\begin{equation}
    \partial_t \rho_{\alpha\beta}(t) = \sum_{\mu\nu} \Lambda_{\alpha\beta,\mu\nu} \rho_{\mu\nu}(t),
    \label{eq:oqs_FBM_modRWA_masterEq}
\end{equation}
where $\rho$ is the density matrix, $\{|\alpha(t)\rangle\}$ are the Floquet states from the driven closed-system Hamiltonian \eqref{eq:driven_hamiltonian}, and $\rho_{\alpha\beta}(t) = \langle \alpha(t)|\rho(t)|\beta(t)\rangle$. The coefficients $\Lambda_{\alpha\beta,\mu\nu}$ are constructed by writing the Born-Markov equation in the Floquet basis (see App. \ref{A:FBM-equation}). They contain all the information from the bath and the system-bath coupling, as well as the driving protocol encoded in the Floquet states. The advantage of working with Eq. \eqref{eq:oqs_FBM_modRWA_masterEq} is that its derivation only assumes standard Born-Markov approximations \cite{kohler_1997,grifoni_1998,hone_2009}, as well as fast drives as compared with the relaxation scale, $\omega_d \tau_\text{rel} \gg1$. These assumptions are naturally satisfied for these fast gates applied to long-$T_1$ qubits. Most importantly, by not restricting the amplitude of the drive, Eq. \eqref{eq:oqs_FBM_modRWA_masterEq} provides the most robust description on the full relevant parameter's regime, while also remaining computationally inexpensive \cite{Mickiewicz2026}.

\begin{figure}[!t]
\includegraphics[width=1.0\linewidth]{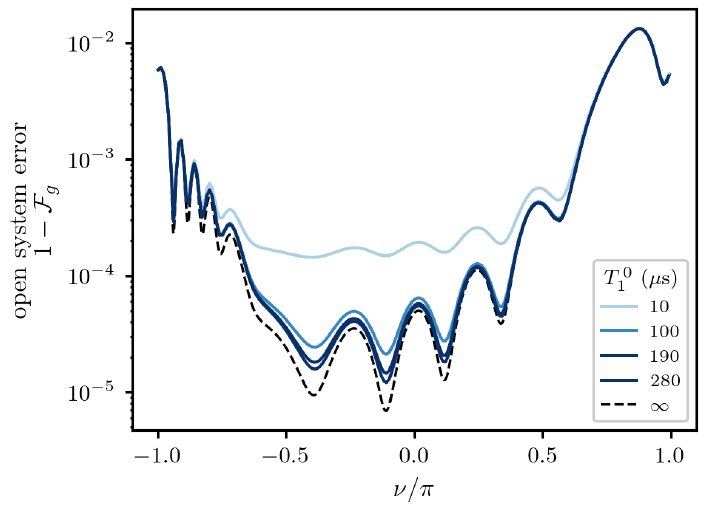}
\caption{\label{fig:fxnm_oqs_error_nu}$Y(\pi/2)$ open-system gate error along the analytical FSW curve, for the fluxonium circuit with parameters $\{ E_J, E_C, E_L\} = \{4,1,1 \}\times 2\pi$ GHz. The noise is modeled as dielectric loss in the capacitor, with an effective bath temperature $T_b=15$ mK, and the different solid lines correspond to different values of $T_1^{\,0}$ (or system-bath coupling). The dashed line shows the closed-system calculation from Sec. \ref{sec:coherent-error}.}
\end{figure}

The model for the system-bath interaction is defined by choosing the noise coupling operator $\hat{\mathcal{O}}$ and setting the coupling constant $\gamma$, the noise power spectrum $S(\omega)$ and the bath temperature $T_b$. We consider a relaxation time $T_1$ limited by dielectric loss, which is believed to be the dominant loss channel near the sweet-spot flux bias, and captures the flux and frequency dependence measured in recent fluxonium experiments \cite{nguyen_2019}. Dielectric loss in fluxonium can be modeled as a lossy shunting capacitor, which couples to the phase $\hat{\varphi}$, and is described by a super-ohmic noise power spectrum of the form
\begin{equation}
    S(\omega)=\gamma\frac{\omega^2}{4E_C}\coth\left(\frac{\omega}{2k_BT_b}\right),
    \label{eq:noise_power_spectrum}
\end{equation}
with $k_B$ the Boltzmann constant. 
$\gamma$ can be estimated from experimental values of $T_1$, reported in literature \cite{nguyen_2019}. These measurements correspond to $T_1$ in the absence of driving, so we note it as $T_1^{\,0}\equiv T_1(A_e\to0)$, but the actual relaxation time $T_1(A_e,\omega_d)$ is strongly affected by the drive amplitude. Using Fermi's golden rule in the approximated two-level system yields $ T_1^{\,0} = \big|\langle 0 | \hat{\mathcal{O}} |1\rangle \big|^2\,S(\omega_q)$, 
thus
\begin{equation}
    \gamma = \frac{4 E_C}{T_1^{\,0}\varphi_{01}^2\omega_q^2} \tanh\left( \frac{\omega_q}{2k_BT_b} \right).
    \label{eq:gamma_oqs}
\end{equation}

For a moderate fluxonium circuit, with relaxation time of $500\,\mu$s, and an effective bath temperature of 15 mK, Eq. \eqref{eq:gamma_oqs} gives $\gamma\sim10^{-7}$. After setting a value of $\gamma$, we extract $T_1(A_e,\omega_d)$, as well as the decoherence time $T_2(A_e,\omega_d)$, by calculating the eigenvalues of the super-operator $\Lambda$ in Eq. \eqref{eq:oqs_FBM_modRWA_masterEq}. We numerically verify that $T_1(A_e\to0,\omega_d)$ converges to the desired $T_1^{\,0}$ value. In the simulations we also find that $T_2(A_e\to0, \omega_d)$ converges to $2 T_1^{\,0}$, which is consistent with a fluxonium device where dephasing is negligible due to the low sensitivity to the noise in the control parameter $\varphi_e$.

Next, we propose a way to measure gate fidelity from Eq. \eqref{eq:oqs_FBM_modRWA_masterEq}. The solution to the master equation is simply given by an exponential of $\Lambda$, $\rho_{\alpha\beta}(t) = \sum_{\mu\nu}(e^{\Lambda t})_{\alpha\beta,\mu\nu} \,\rho_{\mu\nu}(0)$, and transforming back to the $H_0$ eigenbasis, we get $\rho_{ij}(t) = \sum_{kl}\mathcal{U}_{ij,kl}(t) \,\rho_{kl}(0)$, with
\begin{equation}
    \mathcal{U}_{ij,kl}(t)=\sum_{\alpha\beta\mu\nu}c_{i\alpha}(t)c^*_{j\beta}(t)c^*_{k\mu}(0)c_{l\nu}(0) \left(e^{\Lambda t}\right)_{\alpha\beta,\mu\nu},
    \label{eq:superU_ijkl}
\end{equation}
where $c_{i\alpha}(t)\equiv\langle i|\alpha(t)\rangle$. In order to compare $\mathcal{U}(t_g)$ with the process that sends $|\psi\rangle$ into $|\psi '\rangle=U_\text{target} |\psi\rangle$, the appropriate target operator is
\begin{equation}
    \mathcal{U}_\text{target}=U^*_\text{target}\otimes U_\text{target},
    \label{eq:superU_target}
\end{equation}
which is the linear transformation that sends the vectorized density matrix, $|\rho\rangle$, into $|\rho'\rangle=\mathcal{U}_\text{target}|\rho\rangle$. The matrices from Eqs. \eqref{eq:superU_ijkl} and \eqref{eq:superU_target} are equivalent to process matrices obtained in a quantum process tomography (QPT) procedure \cite{Chow2009}. Then, the process fidelity is given by
\begin{equation}
\mathcal{F}_p = \frac{1}{d^2} \text{Tr}\left( \mathcal U^\dagger(t_g)\,\mathcal U_\text{target}\right),
\label{eq:process_fid_oqs}
\end{equation}
and there is a simple relation between $\mathcal F_p$ and the gate fidelity $\mathcal{F}_g$ \cite{Horodecki1999,Nielsen2002},
\begin{equation}
\mathcal{F}_g = \frac{d^2\mathcal F_p + 1}{d^2 + 1},
\label{eq:gate_fid_oqs}
\end{equation}
with $d^2=4$ for the case of single qubit gates.

In Fig. \ref{fig:fxnm_oqs_error_nu} we present the results for $1-\mathcal F_g$, calculated along the analytical FSW curve, with fixed $T_b=15$ mK and different values of $T_1^{\,0}$. Near $\nu=\pm\pi$, where leakage is larger, the coherent error dominates independently of $T_1^{\,0}$. The contrary happens around $\nu=0$, where open-system error is indeed dominated by decoherence processes, and strongly depends on the value of $T_1^{\,0}$. As expected, when $T_1^{\,0}\to\infty$, the open-system calculation converges to the coherent one.

\begin{figure}[!t]
\includegraphics[width=1.0\linewidth]{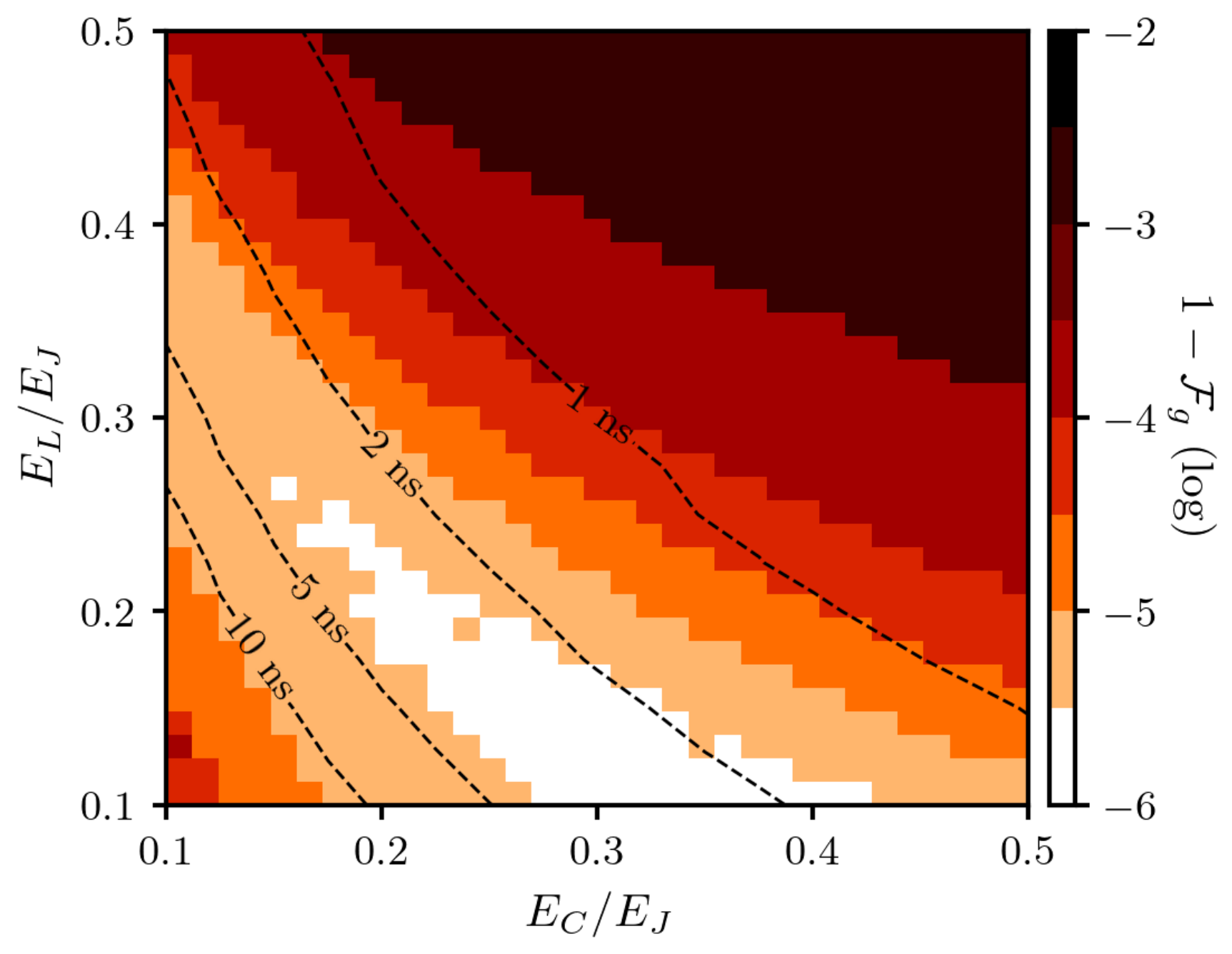}
\caption{\label{fig:fxnm_oqs_sweepELEC}$Y(\pi/2)$ open-system gate error over architecture space within the fluxonium regime, taking the minimum value from $\nu\geq0$. The noise is modeled as dielectric loss in the capacitor, with bath temperature $T_b=15$ mK and $T_1^{\,0}=500\,\mu$s. The dashed contour lines show the total gate time for each architecture, with fixed $E_J/2\pi=4$ GHz.}
\end{figure}

Notably, the open-system error curve still shows deep valleys at the privileged frequencies where leakage is mitigated, even for $T_1^{\,0}$ as low as $10\,\mu$s. This suggests that a similar optimization can be made, sweeping over architecture parameter space and picking the optimal value of $\nu$. Again, we stay with $\nu\geq0$ for faster gate times. In Fig. \ref{fig:fxnm_oqs_sweepELEC} we show a color map of the minimum open-system gate error obtained with a fixed temperature of 15 mK and $T_1^{\,0}=500\,\mu$s as a function of fluxonium parameters. The main distinction from the closed-system calculation shown in Fig. \ref{fig:fxnm_closed_errors}(d) is that we see a much larger error when approaching the heavy-fluxonium regime, mainly due to thermal fluctuations. In that regime, thermal occupation is not negligible since $(1+e^{\omega_q/k_BT_b})^{-1}>1\%$ for qubit frequencies below 200 MHz. It is well known that in this regime reset protocols have to be implemented as well \cite{Najera_2024}. Meanwhile, at the opposite corner of Fig. \ref{fig:fxnm_oqs_sweepELEC}, qubit frequencies approach the GHz zone and anharmonicity is lower. Here, as discussed in Sec. \ref{sec:coherent-error}, leakage becomes the dominating source of error. An optimal architecture region is found around $E_L/E_J\to0.1$, and $E_C/E_J\sim0.5$, where one can achieve an open-system error below the $10^{-5}$ threshold and total gate times lower than $1.3$ ns. For a less conservative threshold of $10^{-4}$, which is sufficient for most of error correction codes \cite{Fowler_2012}, we find some fluxonium architectures which could even overcome the 1 ns gate time.

\section{\label{sec:discussion}Summary and Discussion}



We have analyzed the performance of a strong, one-period LZSM protocol in a superconducting fluxonium qubit, for the implementation of fast and high-fidelity single qubit gates. We derive analytical expressions for the driving parameters $(A_e, \omega_d, t_i,t_f)$, accounting for the multi-level structure of the circuit, which provide accurate seed values that should simplify the experimental calibration work.

We identify that the primary source of error in the closed system is the leakage rate generated by the driving induced transition between the first and second excited states, $|1\rangle\to|2\rangle$. By looking at the frequency spectrum of the pulse, we find the existence of privileged driving frequencies $\omega_d^n=\omega_{21}/n$, with $n\geq2$, at which leakage is suppressed. Operating in these leakage valleys allows for coherent gate errors on the order of $10^{-5}$, with gate times shorter than $2$ ns for a moderated fluxonium architecture. Furthermore, by varying the architecure's parameters, we find that this scheme approaches errors as low as $10^{-10}$ near the heavy-fluxonium regime ($E_C,E_L\ll E_J$), where the qubit's frequency is smaller and anharmonicity is higher. 

The cost of operating in the heavy-fluxonium regime is the longer gate time due to a larger Larmor period $\tau_L=2\pi/\omega_q$, as well as the increasing error contribution from thermal occupation. To analyze the latter, we propose an open-system gate fidelity calculation, $\mathcal{F}_g$, using a Floquet-Born-Markov master equation approach and assuming dielectric loss in the capacitor. This measure allows us to identify an optimal regime around $E_L/E_J\sim0.1$, and $E_C/E_J\sim0.5$, where $1-\mathcal{F}_g < 10^{-5}$ and $t_g<1.3$ ns. Our results show that with these device's parameters quantum control based on LZSM transitions could outperform the resonant driving protocols based on Rabi oscillations.

Interestingly, the control scheme presented here could be extended to two-qubit gates, as shown in the ideal two-level case \cite{Caceres2023}, and demonstrated with similar strong-driving protocols in superconducting qubits \cite{weiss_2022}. We believe that, in order to derive similar analytical tools for the case of two realistic low-frequency qubits, one could aim to a tunable coupling scheme, where the effective coupling should be able to invert the sign with a control parameter. Recent proposals demonstrate the possibility to achieve this, for example by introducing a transmon coupler \cite{Yan2018,moska_2022}. Moreover, a tunable coupling is also promising due to the capability of switching off the interaction, which is believed to be necessary for future scalability \cite{krantz_2019}. Overall, we consider LZSM protocols to be one of the building bricks for implementing quantum computation based on small frequency qubits.

\begin{acknowledgments}
 We acknowledge support from CNEA, ANPCyT (PICT 2019-0654), UNCuyo
(06/C026-T1), CONICET: PIP 11220200101825CO and PIP
11220220100212CO.
\end{acknowledgments}

\appendix

\section{\label{A:CHRW} Counter-rotating hybridized rotating wave approximation}

The underlying two-level system dynamics of the problem has the form of a transversely driven qubit,
\begin{equation}
    H(t)=-\frac{\Delta}{2}\hat{\sigma}_z-\frac{A}{2}\sin(\omega_d t)\hat{\sigma}_x.
    \label{eq:H_TLS_CHRW}
\end{equation}
In the strong driving regime, where $A/\Delta\sim1$, the most typical approaches to solve Eq. \eqref{eq:H_TLS_CHRW}, based on perturbative or secular approximations, fail to capture the effects from the well-known counter-rotating terms \cite{Bloch1940, Shirley1965}. Here we review the results shown in Refs. \cite{Yan2015, Caceres2023}, which preserve these relevant terms by introducing a self-consistent unitary transformation, leading to the so-called counter-rotating hybridized rotating wave approximation (CHRW) \cite{Yan2015}.

First we consider the single pulse with no idles. The time evolution operator after a single-period $T=2\pi/\omega_d$, in this approximation, is given by \cite{Caceres2023}
\begin{equation}
U(T)=
    - \begin{pmatrix}
        a^* & b\\
        -b & a
    \end{pmatrix},
\end{equation}
with
\begin{equation}
    \begin{cases}
        a = \cos(\alpha)+i\sin(\alpha)\cos(\beta),\\
        b = \sin(\alpha)\sin(\beta),
    \end{cases}
\end{equation}
where $(\alpha,\beta)$ depend on the  normalized driving parameters $(A/\Delta,\omega_d/\Delta)$ and a self-consistent parameter $\xi$,
\begin{align}
    &\alpha=\pi\frac{\Delta}{\omega_d}\sqrt{\left( J_0\left( \frac{\xi A}{\omega_d}\right)-\frac{\omega_d}{\Delta}\right)^2 + 4J_1^2\left( \frac{\xi A}{\omega_d}\right)},\\
    &\beta = \arctan\left(\frac{2 J_1\left( \frac{\xi A}{\omega_d}\right)}{J_0\left( \frac{\xi A}{\omega_d}\right)-\frac{\omega_d}{\Delta}}\right) - \frac{\xi A}{\omega_d},\\
    &\xi=1-2\frac{\Delta}{A} J_1\left( \frac{\xi A}{\omega_d}\right),
\end{align}
and $J_n$ is the n-th order Bessel function.

For a $\pi/2$ rotation we set $|b|^2=1/2$, obtaining a non-linear equation for the amplitude and frequency of the pulse,
\begin{equation}
    \sin^2(\alpha)\sin^2(\beta)=\frac12,
    \label{eq:App_CHRW_P01=0.5}
\end{equation}
which results in the closed curve of solutions presented in Sec. \ref{sec:analytical_model}. To be able to perform a specific $\pi/2$ rotation for any of these solutions, we include the idling operations of times $t_i$ and $t_f$,
\begin{equation}
    U(t_g)=e^{i\frac{\Delta}{2}t_f\hat{\sigma}_z}U(T)e^{i\frac{\Delta}{2}t_i\hat{\sigma}_z},
\end{equation}
where $t_g=t_i+T+t_f$ is the total gate time. For any given solution to Eq. \eqref{eq:App_CHRW_P01=0.5}, the final time evolution operator can be rewritten as,
\begin{equation}
    U(t_g)=-\frac{1}{\sqrt 2}\begin{pmatrix}
        e^{i(\varphi_+-\nu)} & e^{i\varphi_-}\\
        -e^{-i\varphi_-} & e^{-i(\varphi_+-\nu)},
    \end{pmatrix}
    \label{eq:App_U(tg)_CHRW}
\end{equation}
where $\varphi_\pm=\Delta(t_f\pm t_i)/2$, and $\nu$ is obtained from
\begin{equation}
    \frac{1}{\sqrt2}e^{i\nu}=\cos(\alpha)+i\sin(\alpha)\cos(\beta).
    \label{eq:App_nu_equation}
\end{equation}
Eq. \eqref{eq:App_nu_equation} turns $\nu$ into a phase that parametrizes the curve of solutions, and in the main degree of freedom in this scheme.

\section{\label{A:idles} From unitary evolution to quantum gates}

From the analytic time evolution operator evaluated at $t_g$, given by Eq. \eqref{eq:App_U(tg)_CHRW}, we calculate the necessary idling times for the implementation of a quantum gate. For simplicity, we pick a target $U_\text{target}=Y(\pi/2)$, where
\begin{equation}
    Y(\pi/2)=e^{-i\frac{\pi}{4}\hat{\sigma}_y}= \frac{1}{\sqrt 2}
    \begin{pmatrix}
        1 & -1\\
        1 & 1
    \end{pmatrix},
    \label{eq:Y_pi/2}
\end{equation}
but the scheme can be generalized for any $\pi/2$ rotation. 

In the case of $A>0$, comparing $U(t_g)$ with $Y(\pi/2)$ gives $\varphi_-=2n\pi$, $\varphi_+=(2m+1)\pi + \nu$, with $n,m\in \mathbb Z$. The shortest-time solution is then,
\begin{equation}
    t_i^{(Y,+A)} = t_f^{(Y,+A)} = \frac{1}{\Delta} \left( \pi + \nu\right),
    \label{A:eq_idles_phi>0}
\end{equation}
resulting in a total idling time $t_i +t_f$ that ranges from 0 to $2\tau_L$, with $\tau_L=2\pi/\Delta$. Note that when $\nu\geq0$, the total idling time is longer than a Larmor period $\tau_L$, which is not desired. An improvement can be achieved by redefining $\nu\to\nu-\pi$, which is actually equivalent to flipping the sign of the amplitude $A\to-A$. This subtle consideration gives a much shorter idling time,
\begin{equation}
    t_i^{(Y,-A)} = t_f^{(Y,-A)} = \frac{\nu}{\Delta}, \quad\text{if }\nu\geq0.
\end{equation}

Therefore, if $\nu<0$ one should operate with $A>0$, and the contrary for $\nu\geq0$. This way, the optimal idles are given by
\begin{equation}
t_i^{Y} = t_f^{Y} =
    \begin{cases}
        (\nu+\pi)/\Delta\,,&\nu\in[-\pi,0),\\
        \nu/\Delta\,,&\nu\in[0,\pi).
    \end{cases}
    \label{eq:idle_times_nu}
\end{equation}
The simple task of picking the optimal sign of the pulse shortens the total gate time by a full $\tau_L$, which becomes significant for low-frequency qubits.


\section{\label{A:FSW-approximation} Floquet-Schrieffer-Wolf effective two-level system}

During the duration of the pulse, the Hamiltonian from Eq. \eqref{eq:driven_hamiltonian} takes the form:
\begin{equation}
    H(t) = H_0 - 2\sin(\omega_d t)\hat{\tilde{\varphi}},
\end{equation}
where $H_0$ is the static fluxonium biased at its sweet-spot, and $\hat{\tilde{\varphi}}\equiv A_e E_L\hat{\varphi}/2$. By construction, the large anharmonicity in $H_0$ makes it so that the Hilbert space can be splitted into two subspaces, one with the low-energy computational basis, and the other with the remaining excited states. This partition naturally sets the ground for Schrieffer-Wolf perturbation theory, which in the case of a time-dependent Hamiltonian needs to be performed in the extended Sambe space by working with the Floquet formalism.

In Ref. \cite{Reparaz2025} a similar problem has been discussed, where $H(t)=H_0+2\cos(\omega_d t)\hat{V}$. Since the cosine and sine functions are almost identical in the Floquet frame, the procedure is analogous and the results can be extended from there. Up to leading order $A_e^2$, the effective Floquet-Schrieffer-Wolf (FSW) two-level system is given by
\begin{equation}
    H_q^\text{FSW}(t) = \tilde{H}_q - 2\tilde{\varphi}_{01}\sin(\omega_d t)\,\hat{\sigma}_x,
\end{equation}
where
\begin{equation}
    \tilde{H}_q = \begin{pmatrix}
        \tilde{E}_0 & \tilde{E}_{01} \\
        \tilde{E}_{01} & \tilde{E}_1
    \end{pmatrix}.
    \label{eq:tildeHq}
\end{equation}
The dressed energies are
\begin{equation}
    \tilde{E}_i = E_i + 2\sum_{j\geq 2} \tilde{\varphi}_{ij}^2 \frac{\omega_{ij}}{\omega_{ij}^2 - \omega_d^2},
    \label{eq:tildeE_i}
\end{equation}
for $i=0,1$, and
\begin{equation}
    \tilde{E}_{01} = \sum_{j\geq 2} \tilde{\varphi}_{0j} \tilde{\varphi}_{1j} \left( \frac{\omega_{0j}}{\omega_{0j}^2 - \omega_d^2} + \frac{\omega_{1j}}{\omega_{1j}^2 - \omega_d^2} \right),
    \label{eq:tildeE_01}
\end{equation}
recalling $\omega_{ij} \equiv E_i - E_j$. Eqs. \eqref{eq:tildeE_i} and \eqref{eq:tildeE_01} are valid for any choice of the flux bias, and in general $\tilde{H}_q$ needs to be re-diagonalized. In the case of operating around the sweet-spot, however, a further simplification can be made by use of the selection rules, where we conveniently obtain $\tilde{\varphi}_{0j} \tilde{\varphi}_{1j}=0\,\forall j$, thus $\tilde{E}_{01}=0$. This reduces the effective two-level system dynamics to a transversely driven qubit with re-normalized frequency,
\begin{equation}
    H_q^\text{FSW}(t) = -\frac12 \tilde{\omega}_q\hat{\sigma}_z - \frac{\tilde{A}}{2}\sin(\omega_d t)\,\hat{\sigma}_x
    \label{},
\end{equation}
with $\tilde{A}=2A_eE_L\varphi_{01}$, and
\begin{equation}
    \begin{aligned}
        &\tilde{\omega}_q(\tilde{A}, \omega_d) = \tilde{E}_1 - \tilde{E}_0\\
        &= \omega_q + \frac{\tilde{A}^2}{2}\sum_{j\geq 2} \left[\varphi_{j0}^2 \frac{\omega_{j0}}{\omega_{j0}^2 - \omega_d^2} - \varphi_{j1}^2 \frac{\omega_{j1}}{\omega_{j1}^2 - \omega_d^2} \right].
    \end{aligned}
    \label{}
\end{equation}

\section{\label{A:fid-optimization} Gate fidelity optimization }

\begin{figure}[!t]
\includegraphics[width=1.0\linewidth]{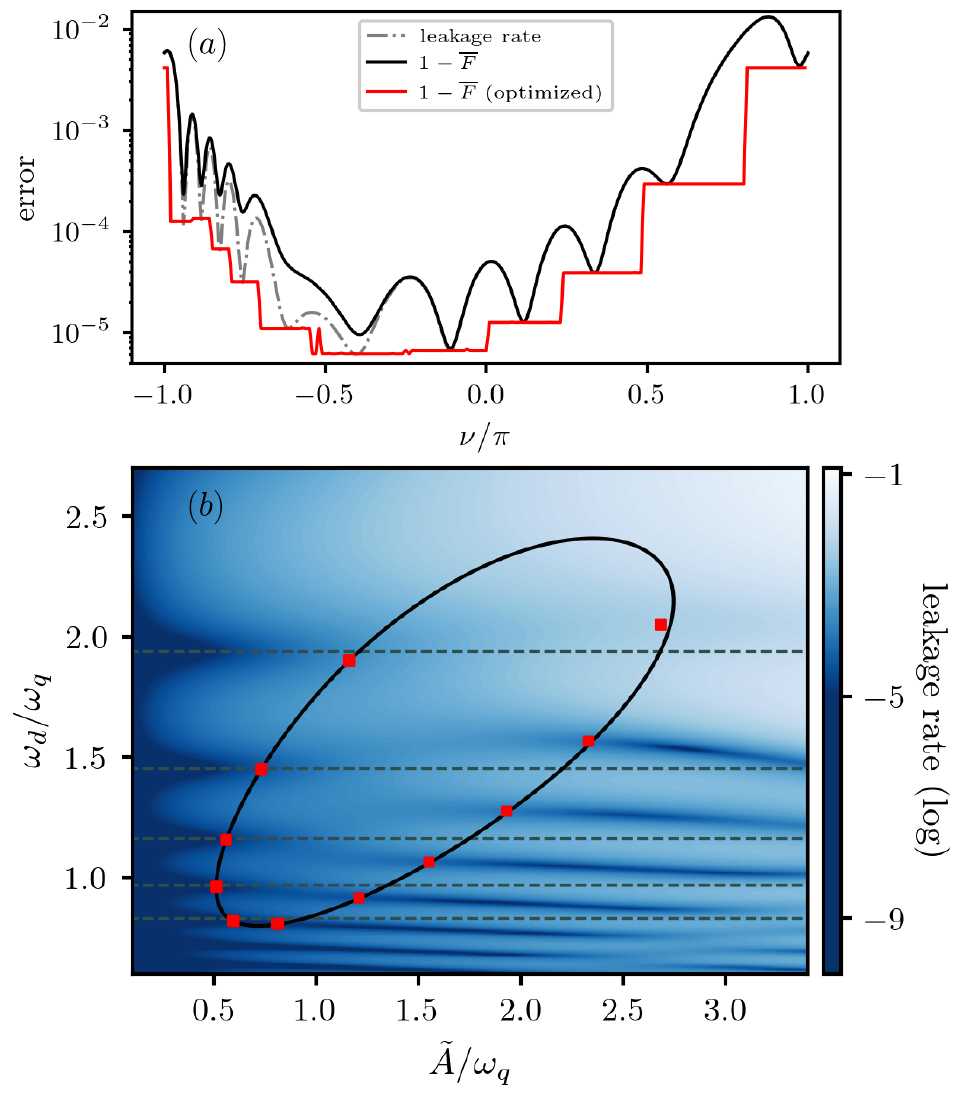}
\caption{\label{fig:fid_optimization}(a) Gate error before and after the optimization (dark and red lines, respectively), for every $\nu$ value on the analytical curve. The dashed-dotted line corresponds to the leakage rate on the analytical curve. (b) Resulting driving parameters after optimization (red dots), compared to the analytical curve (solid line). The colormap shows the leakage rate, verifying that the optimization converges to points over the analytical curve and inside the valleys where leakage is suppressed.}
\end{figure}

Using the analytical driving parameters required to implement a $Y(\pi/2)$ gate, which are parametrized by the phase $\nu$ from  Eq. \eqref{eq:nu_equation}, we first calculate the average gate fidelity $\bar{F}$ and its error $1-\bar{F}$, without any numerical correction. That calculation is presented with a solid dark line in Fig. \ref{fig:fid_optimization}(a), showing that already high fidelity values are obtained from the analytical estimates.

For a further optimization, we take the estimates for each $\nu$ as seed values, and minimize the error numerically using a Nelder-Mead method to search the local minima over the full driving parameter's space. The red line in Fig. \ref{fig:fid_optimization}(a) corresponds to the optimized $1-\bar{F}$ that results from applying the numerical method for each seed. We see that many $\nu$ seeds converge to the same optimal fidelity, forming the discrete flat steps seen in the red curve. The value of the error in each step is exactly equal to the leakage rate at the nearest valley, so one obtains $1-\bar{F}=L$ for the full range of $\nu$. As for the driving parameters, in Fig. \ref{fig:fid_optimization}(b) we show the optimal pairs $(A_e,\omega_d)$, plotted as red dots, and compare them with the analytical FSW curve and the leakage rate colormap. The superposition of both quantities clearly shows that that the optimization method converges to the intersection between the FSW curve and the leakage suppression valleys.


\section{\label{A:leakage_analysis}Leakage rate analysis}

In Sec. \ref{sec:coherent-error} we showed that the gate fidelity in the closed system is ultimately limited by the loss of population outside the computational subspace. Here we present an analytical derivation that lets us estimate the leakage rate present in this scheme.

We start by considering only the lowest 3 energy levels, which according to the definition of leakage in Eq. \eqref{eq:leakage_def} gives
\begin{equation}
    L\approx\frac12\left(|c_{02}|^2+|c_{12}|^2 \right),
    \label{A:eq_leakage_3lvls}
\end{equation}
with $c_{nm}(t) = \langle n|U(t)|m\rangle$. Given the large anharmonicity of the fluxonium spectrum, we can approximate that computational states interact strongly between them but only weakly with the state $|2\rangle$. Therefore, in this approximation we solve (see Ref. \cite{Ferron2010})
\begin{equation}
    i\frac{\partial}{\partial t}c_{n2}(t)=V_{20}(t)e^{i\omega_{20}t}c_{n0}(t) + V_{21}(t)e^{i\omega_{21}t}c_{n1}(t),
\end{equation}
where $V(t)$ is the interaction term from the driven Hamiltonian, namely $V(t)=-2\sin(\omega_d t)\hat{\tilde{\varphi}}$, with $\hat{\tilde{\varphi}}=A_eE_L\hat{\varphi}/2$. Applying the fluxonium selection rules we are left only with $\varphi_{21}\neq0$, resulting in
\begin{equation}
    i\frac{\partial}{\partial t}c_{n2}(t)=-2\tilde{\varphi}_{21}\sin(\omega_d t)e^{i\omega_{21}t}c_{n1}(t).
    \label{A:eq_c2_diffequation}
\end{equation}

Next, we solve Eq. \eqref{A:eq_c2_diffequation} perturbatively by introducing a solution to $c_{n1}(t)$. A standard RWA preformed in the truncated two-level system (with an effective amplitude of $\tilde{A}=2A_eE_L\varphi_{01}$) provides the simplest solution,
\begin{align}
        &c_{01}(t)=e^{i\frac{\omega_d t}{2}}\sin(\theta)\sin\left(\frac{\Omega_r t}{2}\right), \\
        &c_{11}(t)=e^{-i\frac{\omega_d t}{2}}\left(\cos\left(\frac{\Omega_r t}{2}\right) -i\cos(\theta)\sin\left(\frac{\Omega_r t}{2}\right)\right),
\end{align}
with $\tan(\theta)=\tilde{A}/2(\omega_q-\omega_d)$, and the Rabi frequency $\Omega_r=\sqrt{(\omega_q-\omega_d)^2+\tilde{A}^2/4}$. Then, introducing these solutions into Eq. \eqref{A:eq_c2_diffequation} and integrating, we get
\begin{equation}
    c_{02}(t)=-\frac{\tilde{\varphi}_{21}\sin(\theta)}{2}\left( u^{(0)}_{++} + u^{(0)}_{--} - u^{(0)}_{+-} - u^{(0)}_{-+} \right),
    \label{A:eq_c02}
\end{equation}
with $u^{(0)}_{\pm\pm}\equiv e^{i\omega^{(0)}_{\pm\pm}t}/\omega^{(0)}_{\pm\pm}$, and $\omega^{(0)}_{\pm\pm}=\omega_{21}+\omega_d(\frac12 \pm1)\pm\frac{\Omega_r}{2}$, and similarly,
\begin{equation}
\begin{aligned}
    c_{12}(t)=-i\tilde{\varphi}_{21}\Big[&\cos^2\left(\frac{\theta}{2}\right)(u^{(1)}_{+-} - u^{(1)}_{--})\\
    &+ \sin^2\left(\frac{\theta}{2}\right)(u^{(1)}_{++} - u^{(1)}_{-+})\Big],
\end{aligned}
\label{A:eq_c12}
\end{equation}
with $u^{(1)}_{\pm\pm}\equiv e^{i\omega^{(1)}_{\pm\pm}t}/\omega^{(1)}_{\pm\pm}$, and $\omega^{(1)}_{\pm\pm}=\omega_{21}-\omega_d(\frac12 \mp1)\pm\frac{\Omega_r}{2}$.

\begin{figure}[!h]
\includegraphics[width=1.0\linewidth]{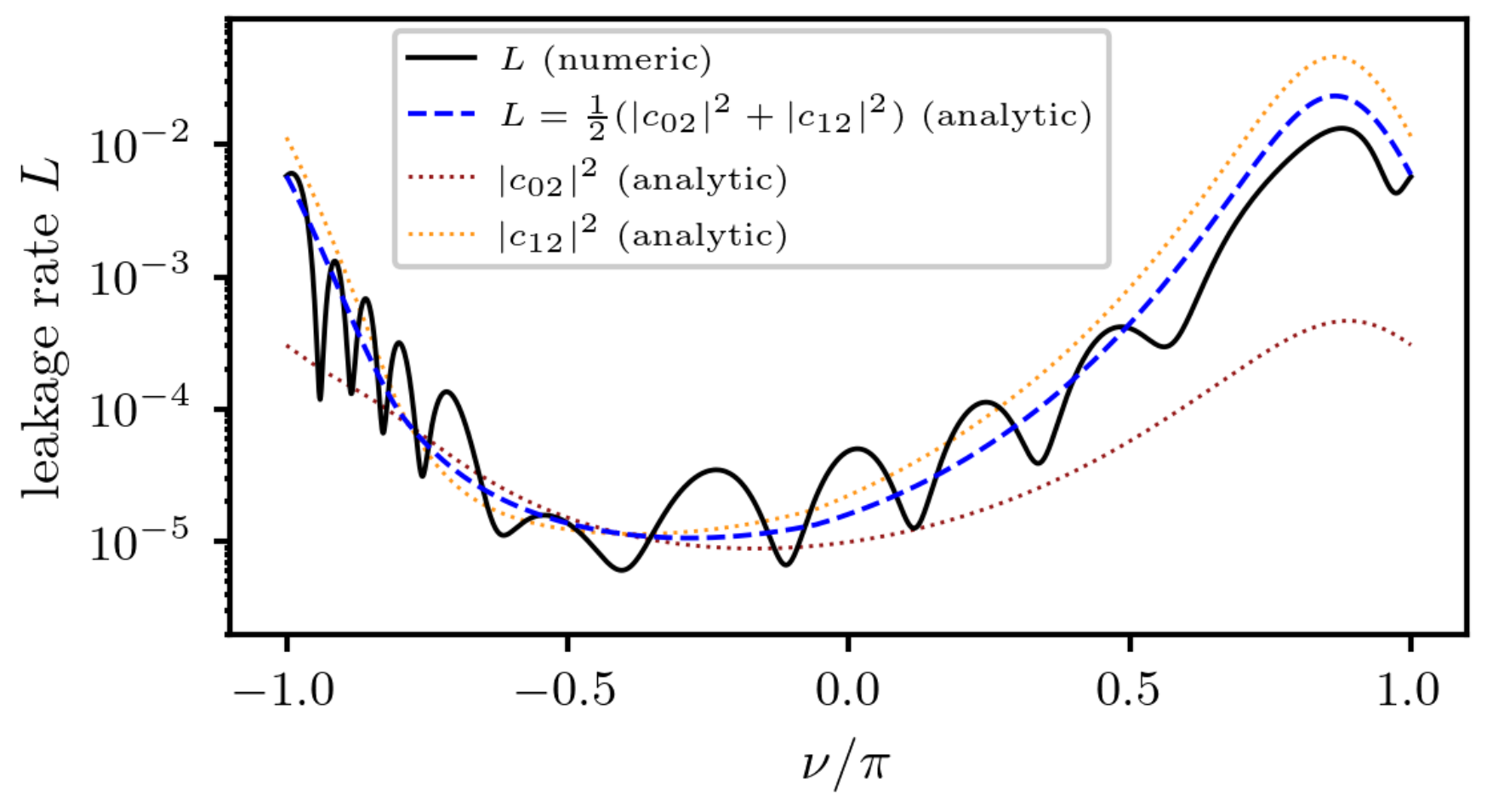}
\caption{\label{fig:leakage_analyitic_nu}Leakage rate $L$ (solid line) over the FSW curve, compared with the analytical calculation from Eq. \eqref{A:eq_leakage_3lvls}, evaluated at the amplitude and frequency corresponding to each $\nu$ (dashed line). The dotted lines show the contribution from each transition.}
\end{figure}

Next, we calculate the leakage rate at the end of the pulse by evaluating Eqs. \eqref{A:eq_c02} and \eqref{A:eq_c12} at $t=2\pi/\omega_d$, and introducing it in Eq. \eqref{A:eq_leakage_3lvls}. In particular, we can calculate $L$ over the FSW curve by plugging in the amplitude and frequency corresponding to each value of $\nu$. In Fig. \ref{fig:leakage_analyitic_nu} we present the resulting $L(\nu)$, compared to the numerically exact calculation. Notably, the analytical expression accurately describes the overall dependency of the curve, and provides a reasonably good order of magnitude. The missing feature, however, is the oscillating behavior that comes from the privileged frequencies described in Sec. \ref{sec:coherent-error}, which seems to be omitted in this approximation.

\begin{figure}[!h]
\includegraphics[width=1.0\linewidth]{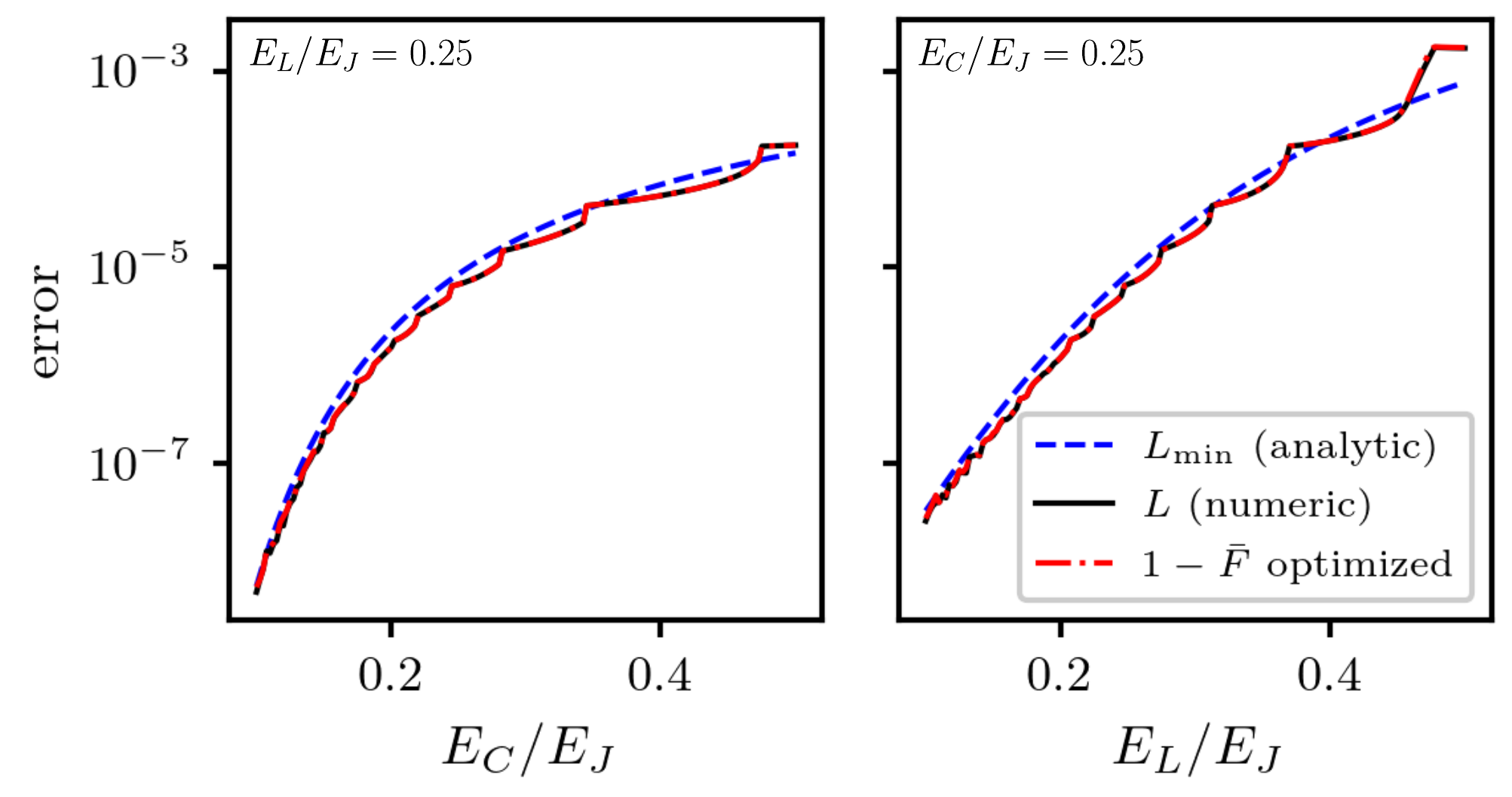}
\caption{\label{fig:leakage_analyitic_ECEL}Analytic $L_\text{min}$ estimation from Eq. \eqref{A:eq_Lmin_analytic} compared with the actual minimum of the leakage rate, as a function of the architecture parameters. The estimation gives a slightly higher error, because it does not account for the privileged frequencies where leakage is mitigated.}
\end{figure}

Finally, from the analytical expression of $L$ we estimate the minimum leakage rate for a given device architecture. To do so, we first note that $L(\nu)$ is minimized around $\nu/\pi\approx-0.25$, which corresponds roughly to $\omega_d\approx\omega_q$, and $\tilde{A}\approx\omega_q/2$. Replacing this values into Eqs. \eqref{A:eq_c02} and \eqref{A:eq_c12}, we get
\begin{equation}
\begin{aligned}
    &|c_{02}|_\text{min} \approx \frac{\varphi_{21}}{4\varphi_{10}} \frac{\omega_q^2 \sin(\pi/4)}{\left(\omega_{21}+\frac12\omega_q \right)^2 - \omega_q^2},\\
    &|c_{12}|_\text{min} \approx \frac{\varphi_{21}}{4\varphi_{10}} \frac{\omega_q^2 \cos(\pi/4)}{\left(\omega_{21}+\frac12\omega_q \right)^2 - \omega_q^2}.
\end{aligned} 
\end{equation}
Therefore, we arrive at a simple analytic expression to estimate the minimum leakage rate,
\begin{equation}
    L_\text{min} \sim \frac{1}{32}\left(\frac{\varphi_{21}}{\varphi_{10}}\right)^2\frac{\omega_q^4}{\left( (\omega_{21} + \frac12\omega_q)^2-\omega_q^2 \right)^2}.
    \label{A:eq_Lmin_analytic}
\end{equation}
In Fig. \ref{fig:leakage_analyitic_ECEL} we show the result of calculating $L_\text{min}$ as a function of the device's parameters. Comparing the analytical estimation from Eq. \eqref{A:eq_Lmin_analytic} with the actual minimum leakage rate, we see a surprisingly good agreement, although the exact value is slightly smaller because of the leakage suppression valleys.

Note that from $L_\text{min}$ we can also estimate with good accuracy the order of magnitude of the maximum obtainable fidelity, $\bar{F} \sim 1 - L_\text{min}$. Furthermore, we can do a final simplification of Eq. \eqref{A:eq_Lmin_analytic} in the case of $\omega_{21}\gg\omega_q$, or large anharmonicity $\alpha = \omega_{21}-\omega_q$. This way we obtain a rule of thumb equation for the fidelity,
\begin{equation}
    1 - \bar{F} \sim \frac{1}{32}\left(\frac{\varphi_{21}}{\varphi_{10}}\right)^2\left(1 - \frac{\alpha}{\omega_{21}}\right)^4,
\end{equation}
which could be handful for device engineer.

\section{\label{A:FBM-equation} Floquet-Born-Markov master equation}

We model the open-system dynamics from the total Hamiltonian,
\begin{equation}
    H_\text{total}(t)=H_s(t) + H_b + H_{sb},
\end{equation}
where $H_s(t)$ is the driven fluxonium Hamiltonian given by Eq. \eqref{eq:driven_hamiltonian}, and $H_b=\sum_k\omega_kb^\dagger_kb_k$ is the Hamiltonian of a thermal bosonic bath at temperature $T_b$. The coupling is given by $H_{sb}=\mathcal{O}\otimes\mathcal{B}$, with $\mathcal{O}$ and $\mathcal{B}$ operators from the system and the bath, respectively.

Since we have the freedom to extend $H_s(t)$ arbitrarily to $t\to\pm\infty$, a convenient choice is to extend it periodically, with a period given by $\tau=t_i+2\pi/\omega_d+t_f$. The Hamiltonian now satisfies $H_s(t) = H_s(t+\tau)$, and we can invoke the Floquet formalism to define a basis of solutions of the form $|\psi_\alpha(t)\rangle=\exp(-i\varepsilon_\alpha t)|\alpha(t)\rangle$, with $|\alpha(t)\rangle = |\alpha(t+\tau)\rangle$. The Floquet states $|\alpha(t)\rangle$ and the corresponding quasienergies, $\varepsilon_\alpha$, are obtained numerically by diagonalizing the time-evolution operator after one period, $U(t+\tau,t)=\sum_\alpha \exp(-i\varepsilon_\alpha \tau)|\alpha(t)\rangle\langle\alpha(t)|$.

Performing the Born-Markov approximations, and moving into the Floquet frame, one obtains the Floquet-Born-Markov (FBM) master equation \cite{kohler_1997,grifoni_1998,hone_2009} for the density matrix of the closed system $\rho(t)\equiv\rho_s(t)$,
\begin{equation}
    \partial_t \rho_{\alpha\beta}(t) = \sum_{\mu\nu} \Lambda_{\alpha\beta,\mu\nu}(t) \rho_{\mu\nu}(t),
    \label{eq:FBM_timedependent}
\end{equation}
where $\rho_{\alpha\beta}(t) \equiv \langle\alpha(t)|\rho(t)|\beta(t)\rangle$, and $\Lambda(t)$ is a Liouvillian super-operator, whose coefficients are given by
\begin{equation}
    \Lambda_{\alpha\beta,\mu\nu}(t) = -i(\varepsilon_\alpha - \varepsilon_\beta)\,\delta_{\alpha\mu}\delta_{\beta\nu} + R_{\alpha\beta,\mu\nu}(t).
\end{equation}
The rates $R_{\alpha\beta,\mu\nu}(t)$ are periodic in time, so can be Fourier decomposed,
\begin{equation}
    R_{\alpha\beta,\mu\nu}(t) = \sum_{k,q}e^{i(k+q)\Omega t} R^{k,q}_{\alpha\beta,\mu\nu},
    \label{eq:coeficientes_R_t}
\end{equation}
with $\Omega=2\pi/\tau$, and
\begin{equation}
\begin{aligned}
        R^{k,q}_{\alpha\beta,\mu\nu} = \,&\mathcal{O}^k_{\alpha\mu} \mathcal{O}^q_{\nu\beta}\left[ g_+(-\varepsilon^k_{\alpha\mu}) + g_-(-\varepsilon^q_{\nu\beta})\right] \\
        &- \delta_{\nu\beta}  \sum_\lambda\mathcal{O}^k_{\alpha\lambda} \mathcal{O}^q_{\lambda\mu} g_+(-\varepsilon^q_{\lambda\mu})\\
        &- \delta_{\mu\alpha}  \sum_\lambda\mathcal{O}^q_{\nu\lambda} \mathcal{O}^k_{\lambda\beta} g_-(-\varepsilon^k_{\nu\lambda}),
\end{aligned}
\label{eq:rates_R_FBM}
\end{equation}
where $\mathcal{O}^k_{\alpha\beta}=\frac{1}{\tau}\int_0^\tau dt\,e^{-ik\Omega t} \langle\alpha(t)|\mathcal{O}|\beta(t)\rangle$, $g_\pm(\omega)$ are the spectral bath correlation functions, and $\varepsilon^k_{\alpha\beta} = \varepsilon_\alpha-\varepsilon_\beta+k\Omega$. The functions $g_\pm(\omega)$ are typically redefined in terms of the spectral density of the bath, $J(\omega)$,
\begin{equation}
    g_+(\omega) + g_-(\omega)=J(\omega)\left( n_{T_b}(\omega)+1\right),
\end{equation}
where $n_{T_b}(\omega)=(\exp(\omega/k_BT_b)-1)^{-1}$. In terms of the noise power spectrum, $S(\omega)=J(\omega)\coth(\omega/2k_BT_b)$.

Note that Eq. \eqref{eq:FBM_timedependent} has the form of a Redfield master equation, with the important difference that $\Lambda(t)$ is a sum of many time-dependent terms, making it numerically expensive. In order to obtain a time-independent $\Lambda$, one can approximate the rates $R(t)$ by its average \cite{kohler_1997,grifoni_1998,hone_2009}, by neglecting the oscillating terms in Eq. \eqref{eq:coeficientes_R_t} corresponding to $k\neq-q$. This is valid whenever the relaxation scale $\tau_\text{rel}$ is much larger than $\tau$, or roughly $\omega_d\tau_\text{rel}\gg 1$. Crucially, the remaining rates preserve the relevant dependency with the amplitude and frequency of the drive, and we get $\Lambda_{\alpha\beta,\mu\nu}(t) \approx \Lambda_{\alpha\beta,\mu\nu}$, with
\begin{equation}
    \Lambda_{\alpha\beta,\mu\nu} = -i(\varepsilon_\alpha - \varepsilon_\beta)\,\delta_{\alpha\mu}\delta_{\beta\nu} + \sum_k R^{k,-k}_{\alpha\beta,\mu\nu}.
\end{equation}
This step is often called as a moderated rotating-wave approximation, since the argument is similar to a RWA but the hypothesis is weaker. In fact, one can transform Eq. \eqref{eq:FBM_timedependent} into a Lindbladian form by applying a second secular approximation in the interaction picture, but this is not desired due to the incremented error appearing at the quasi-energies crossings \cite{Mickiewicz2026}. Therefore, we stay with the moderated FBM equation,
\begin{equation}
    \partial_t \rho_{\alpha\beta}(t) = \sum_{\mu\nu} \Lambda_{\alpha\beta,\mu\nu}\, \rho_{\mu\nu}(t),
    \label{eq:FBM_timeindependent}
\end{equation}
which provides a robust overall description while remaining computationally inexpensive.


\bibliography{apssamp}

\end{document}